\documentclass[12pt]{article}

\title{\bf Quantum-theoretical description of the cosmological
constant as the zero-point energy}
\author{BUDH RAM \\
Physics Department \\
New Mexico State University\\ 
Las Cruces, New Mexico 88003, USA\\ 
 \\
 and \\
 \\
 Prabhu-Umrao Institute of Fundamental Research \\
 A2/214 Janak Puri, New Delhi, 110058, India \\
}

\begin{document}

\maketitle

\begin{abstract} 
The present paper seeks to construct a quantum theory of the
cosmological constant in which its presently observed very small value
emerges naturally.
\end{abstract}

Large scale cosmological observations over the past decade (see
references cited in [1], and [2-4]) have provided data which are in
accord with the simplest cosmological model LCDM ($\Lambda$ + Cold Dark
Matter) [4-7].  This model is governed by the homogeneous and
isotropic Friedmann-Robertson-Walker (FRW) equation
\begin{equation}
\left({\dot a \over a}\right)^2 = {8\pi \over 3} \rho_M + {\Lambda
\over 3}
\label{one}
\end{equation}
for the scale factor $a(t)$,
with $\Omega_\Lambda + \Omega_M = 1$ (flat universe). (Units in which $G
= \hbar = c = 1$ are used throughout this paper.) Eq. (1), with
$\rho_M \propto a(t)^{-3}$, has the analytical solution [8-10]
\begin{equation}
a(t) \propto \left(\sinh {3\over2} \sqrt{\Lambda \over 3}
~t\right)^{2/3}.
\label{two}
\end{equation}
The cosmological data are well fitted by
the LCDM with $\Omega_\Lambda = {\Lambda \over 3H^2_0} \simeq 0.7$ and
$h \simeq 0.7$ [4-6], where $H_0$ (the present value of the Hubble
parameter) = 100 $h$ km s$^{-1}$ Mpc$^{-1}$.

The big question, then, is : How can one account for the very small
value $1.02 \times 10^{-61}$ of $\sqrt{\Lambda \over 3}$ --- obtained
from the above values of $\Omega_\Lambda$ and $h$ --- theoretically
[10-14]? Or, put more concretely, can one construct a
quantum-mechanical description in which this small value of
$\sqrt{\Lambda \over 3}$ emerges naturally?  The answer is in the
affirmative.  In the following I wish to sketch briefly the new
quantum construction.

I start with the Einstein equation
\begin{equation}
R_{\mu\nu} - {1\over2} g_{\mu\nu} R = \Lambda g_{\mu\nu}
\label{three}
\end{equation}
and its solution [15,16] --- the de Sitter line element ---
\begin{equation}
ds^2 = (1 - \omega^2 \rho^2) dt^2 - {d\rho^2 \over (1 - \omega^2
\rho^2)} - \rho^2 (d\theta^2 + \sin^2\theta d\phi^2),
\label{four}
\end{equation}
or, equivalently,
\begin{eqnarray}
ds^2 = 4\pi \Bigg[{1\over 4\pi} (1 - \omega^2 4\pi r^2) dt^2 &-& {dr^2
\over (1 - \omega^2 4\pi r^2)} - \nonumber \\[2mm] 
&-& r^2 (d\theta^2 + \sin^2 \theta d\phi^2)\Bigg]
\label{five}
\end{eqnarray}
by setting $\rho/\sqrt{4\pi} = r$.  In Eqs. (4) and (5), $\omega =
\sqrt{\Lambda / 3} = \sqrt{8\pi\rho_v / 3}$.
The de Sitter line element represents a completely homogeneous
space-time manifold. Note that the metric (5) is similar to the
Schwarzschild metric 
\begin{equation}
ds^2 = \left(1 - {2M \over r}\right) dt^2 - {dr^2 \over \left(1 - {2M
\over r}\right)} - r^2 (d\theta^2 + \sin^2 \theta d\phi^2).
\label{six}
\end{equation}
Just as the Schwarzschild metric has a horizon at $r = 2M$, the de
Sitter metric has a horizon at $4\pi r^2 = {1 \over \omega^2}$.
(Note that ${1 \over \omega^2}$ is the surface area of the
horizon.) In order to construct quantum mechanics for the de Sitter
line element (5) I use the same method which is used in [17] for
making quantum mechanics for the Schwarzschild line element (6).
There is one difference.  In the Schwarzschild case it is the
time-like geodesic equation that is used, for the simple reason that
$t$ is time-like for $r > 2M$ [18,19].  In the present de Sitter case,
however, I have to use the {\it space-like} geodesic equation because
in this case $t$ is {\it space-like} for $4\pi r^2 > {1 \over
\omega^2}$.

The equation for the space-like geodesic with the de Sitter line
element (5) can be easily obtained [20], and is given by
\begin{equation}
- {1 \over 4\pi} = \left[{E^2 \over (1 - 4\pi \omega^2 r^2)} - {{\dot
r}^2 \over (1 - 4\pi \omega^2 r^2)} - {L^2 \over r^2}\right]
\label{seven}
\end{equation}
with $E$ and $L$ constants of integration.  In Eq. (7) dot means
differentiation with respect to the proper time.  Following [17] I put
$E = L = 0$ in (7), and obtain
\begin{equation}
{1\over2} {\dot r}^2 + {1 \over 2} \omega^2 r^2 = {1 \over 8\pi}.
\label{eight}
\end{equation}
Proper use of the Schr\"odinger prescription in (8) then leads to
\begin{equation}
\left[- {1 \over 2r^2} {\partial \over \partial r} \left(r^2 {\partial
\over \partial r}\right) + {1 \over 2} \omega^2 r^2\right] \psi = {1
\over 8\pi} \psi
\label{nine}
\end{equation}
as the quantum equation that corresponds to the Einstein equation (3)
with the classical de Sitter solution (5).  With $U = r \psi$, Eq. (9)
reduces to the simple form
\begin{equation}
\left(-{1 \over 2} {d^2 \over dr^2} + {1 \over 2} \omega^2 r^2\right)
U = {1 \over 8\pi} U.
\label{ten}
\end{equation}

Eq. (10) is the three-dimensional quantum oscillator but with only one
eigenvalue $\epsilon = {1 \over 8\pi}$ in
\begin{equation}
\epsilon = \left(2n + {3 \over 2}\right)\omega, \ n = 0,1,2,\cdots
\label{eleven}
\end{equation}
see [17].
It is most natural to assign this value to the $n = 0$ ground state --
the state of {\it zero-point energy}:
\begin{equation}
{1 \over 8\pi} = {3 \over 2} \omega
\label{twelve}
\end{equation}
which gives $\omega = {1 \over 12\pi}$, or writing $\omega$ explicitly
in terms of $\Lambda$, $\sqrt{\Lambda \over 3} = {1 \over 12\pi}$.
Denoting this value of $\Lambda$ by $\Lambda_0$ and the corresponding
$\omega$ by $\omega_0$, one has $\omega_0 = \sqrt{\Lambda_0 \over 3} =
{1 \over 12\pi}$, and (12) is rewritten as
$$
{3 \over 2} \omega_0 = {1 \over 8\pi}.
\eqno (12')
$$
On the other hand, if one associates this energy eigenvalue ${1 \over
8\pi}$ to the $n = 1$ state, then from (11) 
\begin{equation}
{1 \over 8\pi} \equiv {3 \over 2} \omega_0 = \left(2(1) + {3 \over
2}\right) \omega,
\label{thirteen}
\end{equation}
which gives $\omega = {3\omega_0 \over 7} = {1 \over 28\pi}$.
Denoting this value of $\omega$ by $\omega_1$, and the corresponding
value of $\Lambda$ by $\Lambda_1$, one has $\omega_1 = \sqrt{\Lambda_1
\over 3} = {1 \over 28\pi}$, and ${3\over2} \omega_1 = {3 \over
56\pi}$.  And Eq. (13) is rewritten as
$$
{3 \over 2} \omega_0 = \left(2(1) + {3\over2}\right) \omega_1.
\eqno(13')
$$
Eq. ($13^\prime$) has the obvious interpretation in view of [17,19], that the
initial zero-point energy ${3\over2} \omega_0$ is reduced to
${3\over2} \omega_1$, the difference ${3\over2} (\omega_0 - \omega_1)
= 2\omega_1$ being converted to one pair of mass quanta (particles),
each mass quantum of energy $\omega_1$.  Said another more physical
way, the first ``quantum bang'' (QB) transforms an amount $2\omega_1$
of the zero-point energy into (dark) matter.

Next associate the left-over zero-point energy ${3\over2} \omega_1$ to
the eigenvalue $\epsilon$ for the $n = 2$ state in (11):
\begin{equation}
{3\over2} \omega_1 = \left(2(2) + {3\over2}\right) \omega_2,
\label{fourteen}
\end{equation}
giving $\omega_2 = {3\omega_1 \over 4(2) + 3}$ and the further reduced
value of the zero-point energy $= {3\over2} \omega_2$, the difference
${3\over2} (\omega_1 - \omega_2) = 4\omega_2$ being converted to 2
pairs of mass quanta (particles), each particle of energy $\omega_2$.
Again said more physically, the second successive QB transforms
further an amount $4\omega_2$ of the zero-point energy into (dark)
matter. 

Similarly, after associating the energy ${3\over2} \omega_2$ with the
eigenvalue of the $n = 3$ state, one obtains
\begin{equation}
{3\over2} \omega_2 = \left(2(3) + {3\over2}\right) \omega_3,
\label{fifteen}
\end{equation}
with the zero-point energy reduced further to the value ${3\over2}
\omega_3$ and $3$ pairs of particles, each of energy $\omega_3$,
produced. 

Continuing the process, after a succession of $n$ QB$^\prime$s,
one gets
\begin{equation}
{3\over2} \omega_{n-1} = \left(2n + {3\over2}\right) \omega_n,
\label{sixteen}
\end{equation}
which gives the recursion relation
\begin{equation}
\omega_n = {3\omega_{n-1} \over 4n+3}
\label{seventeen}
\end{equation}
from which numerical values of $\omega_n ^\prime$s can be easily obtained
starting from $\omega_0 = {1 \over 12\pi}$.  To give a feel for how
they decrease with increasing $n$, I list a few : $\omega_0 = {1 \over
12\pi}$, $\omega_1 = {1 \over 28\pi}$, $\omega_2 = 3.10 \times
10^{-3}$, $\cdots$, $\omega_{14} = 6.58 \times 10^{-16}$, $\omega_{15}
= 3.13 \times 10^{-17}$, $\omega_{16} = 1.40 \times 10^{-18}$,
$\cdots$, $\omega_{28} = 2.04 \times 10^{-36}$, $\omega_{29} = 5.14
\times 10^{-38}$, $\omega_{30} = 1.25 \times 10^{-39}$, $\cdots$,
$\omega_{42} = 5.86 \times 10^{-60}$, $\omega_{43} = 1.00 \times
10^{-61}$, $\omega_{44} = 1.68 \times 10^{-63}$, $\cdots$.

The value $\omega_{43} = \sqrt{\Lambda_{43} \over 3} = {\bf 1.00 \times
10^{-61}}$ gives $(\rho_v)^{1/4} = 1.86 \times 10^{-31} [= {\bf 2.27
\times 10^{-3} {\rm\bf eV}}$] via the relation $8\pi\rho_v =
\Lambda$; and with $\Omega_\Lambda = 0.7$, it gives $h = {\bf 0.69}$.  

The classical solution (2) has
${3\over2} \sqrt{\Lambda \over 3}$ in the argument of the
function $\sinh$.  The quantum zero-point energy --- 
${3\over2} \sqrt{\Lambda \over 3}$ ---
has kept its {\it form}
--- in the classical FRW solution ---
[21], though at a much reduced value equal to $1.5 \times 10^{-61}$.
Thus the classical description (2) is a continuation of the above
quantum description, thereby authenticating both, the former taking
over the latter after most of the zero-point energy has been converted
into dark and undark matter, in strict adherence with the law of
conservation of energy, as a result of the Quantum Big Bang -- 
a succession of forty three quantum bangs. Captivatingly simple.

\bigskip

\centerline{\bf Acknowledgments}
\medskip

I thank Arun Ram, Nilam Ram and Rajeev Bhalerao for critical reading
of the manuscript, Jim Shirley for doing the numerical calculation,
and Naresh Dadhich for an interesting conversation.  Thanks are due
Narayan Banerjee for suggesting that I apply the method [17] to the
early universe, for sending me the references to review articles, and
for many helpful telephone conversations.  It is a pleasure to
acknowledge receipt of an e-mail from Samuel Braunstein in which he
asked a question which aroused my curiosity to see what would happen
if I applied the method [17] to the de Sitter line element.

\bigskip
\bigskip

\centerline{\bf References and Notes}
\medskip

\begin{enumerate}
\item[{[1]}] E. Copeland, M. Sami, and S. Tsujikawa, ArXiv:
hep-th/0603057 (2006).
\item[{[2]}] A.G. Riess {\it et al}, Ap J. {\bf 627}, 607 (2005).
\item[{[3]}] A.G. Riess {\it et al}, ArXiv: astro-ph/0611572 (2006).
\item[{[4]}] D.N. Spergel {\it et al}, ArXiv: astro-ph/0603449v2
(2007). 
\item[{[5]}] L. Perivolaropoulos, ArXiv: astro-ph/0601014 (2006).
\item[{[6]}] A.G. Riess {\it et al}, Ap J. {\bf 607}, 665 (2004).
\item[{[7]}] M.S. Turner and A.G. Riess, Ap J. {\bf 569}, 18 (2002).
\item[{[8]}] H. Bondi, {\it Cosmology} (Cambridge Univ. Press, 1960). 
\item[{[9]}] P.J.E. Peebles, Ap J. {\bf 284}, 439 (1984).
\item[{[10]}] V. Sahni and A. Starobinsky, Int. J. Mod. Phys. 
{\bf D9}, 373 (2000).      
\item[{[11]}] S.M. Carroll, ArXiv: astro-ph/0310342 (2003).
\item[{[12]}] S. Weinberg, Rev. Mod. Phys. {\bf 61}, 1 (1989).
\item[{[13]}] The presence of $\Lambda$ in the Einstein equations is
equivalent to the presence of the energy density and pressure of a
fluid ($\Lambda$ fluid) or simply of vacuum [14] with $\rho_\Lambda
\equiv \rho_v = {\Lambda \over 8\pi}$; $p_\Lambda \equiv p_v = -{\Lambda
\over 8\pi}$.  In present day particle physics this energy density
$\rho_v$ is associated with the zero-point energy of the quantum
field(s) that permeate the {\it vacuum}.  A simple calculation,
however, of this quantity even for a single bosonic field gives an
infinite value [10].  Consequently theorists have made guesses; and
the answer depends on the manner in which one makes a guess.  The
upshot is that none of those guesses has lead to anything concrete
[10,11,12]. 
\item[{[14]}] Ya.B. Zel$^\prime$dovich, Sov. Phys. Uspekhi {\bf 11}, 381 (1968).
\item[{[15]}] H.P. Robertson, Phil. Mag. {\bf 5}, 835 (1928).
\item[{[16]}] T. Padmanabhan, Phys. Rep. {\bf 380}, 235 (2003).
\item[{[17]}] B. Ram, A. Ram, and N. Ram, ArXiv: gr-qc/0504030 (2005);
B. Ram, Phys. Lett. {\bf A265}, 1 (2000).
\item[{[18]}] Likewise the time-like geodesic equation is used in
[19], as in the Ba\~nados, Teitelboim, Zanelli (BTZ) line-element
also, $t$ is time-like for $r > \sqrt{8M} \ell$.     
\item[{[19]}] B. Ram and J. Shirley, ArXiv: gr-qc/0604074 (2006).
\item[{[20]}] By following the method given on pages 96-97 of
S. Chandrasekhar, {\it The Mathematical Theory of Black Holes},
Oxford Univ. Press 1983.  
\item[{[21]}] ``This is beautiful'', said my friend Rebecca
spontaneously when I pointed this out to her. 
\end{enumerate}
  
\end{document}